\documentclass[aps,prd,twocolumn,groupedaddress,amssymb,eqsecnum,showpacs,epsfig,nofootinbib]{revtex4}
\usepackage{graphicx}
\usepackage{bm}
\usepackage{dcolumn}
\usepackage{amsmath}
\usepackage{amsmath}
\usepackage{amssymb}

\numberwithin{equation}{section}
\usepackage{epsfig}
\def \lleq {\lower0.9ex\hbox{ $\buildrel < \over \sim$} ~}
\def \ggeq {\lower0.9ex\hbox{ $\buildrel > \over \sim$} ~}

\def \om    {\Omega}
\newcommand{\ben}{\begin{eqnarray}}
\newcommand{\een}{\end{eqnarray}}

\def \obh {\Omega_b h^2}
\def \ob  {\Omega_b}

\def \omms  {\Omega_m}

\def \omm  {\Omega_{0 {\rm m}}}
\def \om  {\Omega_m}

\def \beq  {\begin{equation}}
\def \eeq  {\end{equation}}
\def \ber  {\begin{eqnarray}}
\def \eer  {\end{eqnarray}}

\newcommand {\ga} {\ {\raise-.5ex\hbox{$\buildrel>\over\sim$}}\ }
\newcommand {\la} {\ {\raise-.5ex\hbox{$\buildrel<\over\sim$}}\ }

\begin{document}
\newcommand{\newc}{\newcommand}
\newc{\be}{\begin{equation}}
\newc{\ee}{\end{equation}}
\newc{\ba}{\begin{eqnarray}}
\newc{\ea}{\end{eqnarray}}
\newc{\bea}{\begin{eqnarray*}}
\newc{\eea}{\end{eqnarray*}}
\newc{\D}{\partial}
\newc{\ie}{{\it i.e.} }
\newc{\eg}{{\it e.g.} }
\newc{\etc}{{\it etc.} }
\newc{\etal}{{\it et al.}}
\newc{\lcdm }{$\Lambda$CDM }
\newcommand{\nn}{\nonumber}
\newc{\ra}{\rightarrow}
\newc{\lra}{\leftrightarrow}
\newc{\lsim}{\buildrel{<}\over{\sim}}
\newc{\gsim}{\buildrel{>}\over{\sim}}
\title{Comparison of recent SnIa datasets}
\author{J. C. Bueno Sanchez$^a$, S. Nesseris$^b$ and L. Perivolaropoulos$^a$}
\affiliation{$^a$Department of Physics, University of Ioannina,
Greece\\ $^b$ The Niels Bohr International Academy, The Niels Bohr
Institute, Blegdamsvej 17, DK-2100, Copenhagen \O, Denmark}
\date{\today}

\begin{abstract}
We rank the six latest Type Ia supernova (SnIa) datasets
(Constitution (C), Union (U), ESSENCE (Davis)  (E), Gold06 (G),
SNLS 1yr (S) and SDSS-II (D)) in the context of the
Chevalier-Polarski-Linder (CPL) parametrization $w(a)=w_0+w_1
(1-a)$, according to their Figure of Merit (FoM), their
consistency with the cosmological constant ($\Lambda$CDM), their
consistency with standard rulers (Cosmic Microwave Background
(CMB) and Baryon Acoustic Oscillations (BAO)) and their mutual
consistency. We find a significant improvement of the FoM (defined
as the inverse area of the $95.4\%$ parameter contour) with the
number of SnIa of these datasets ((C) highest FoM, (U), (G), (D),
(E), (S) lowest FoM). Standard rulers (CMB+BAO) have a better FoM
by about a factor of 3, compared to the highest FoM SnIa dataset
(C). We also find that the ranking sequence based on consistency
with $\Lambda$CDM is identical with the corresponding ranking
based on consistency with standard rulers ((S) most consistent,
(D), (C), (E), (U), (G) least consistent). The ranking sequence of
the datasets however changes when we consider the consistency with
an expansion history corresponding to evolving dark energy
$(w_0,w_1)=(-1.4,2)$ crossing the phantom divide line $w=-1$ (it
is practically reversed to (G), (U), (E), (S), (D), (C)). The
SALT2 and MLCS2k2 fitters are also compared and some peculiar
features of the SDSS-II dataset when standardized with the MLCS2k2
fitter are pointed out. Finally, we construct a statistic to
estimate the internal consistency of a collection of SnIa
datasets. We find that even though there is good consistency among
most samples taken from the above datasets, this consistency
decreases significantly when the Gold06 (G) dataset is included in
the sample.
\end{abstract}
\maketitle

\section{Introduction}
The accelerating expansion of the universe has been indicated
consistently by a wide range of cosmological data. The most
sensitive probes of this expansion currently are standard candles
in the form of Type Ia supernovae (SnIa) \cite{lr,scp,hzsst,Riess:2004nr,Hicken:2009dk,Kowalski:2008ez,Davis:2007na,Astier:2005qq,Riess:2006fw,Kessler:2009ys} and standard rulers in the form
of the sound horizon scale at last scattering as measured through
the Cosmic Microwave Background (CMB) power spectrum
\cite{Spergel:2006hy,Komatsu:2008hk} and through Baryon Acoustic
Oscillations (BAO)
\cite{Eisenstein:2005su,Percival:2007yw,Percival:2009xn}. These
cosmological observations have
indicated\cite{Spergel:2006hy,Komatsu:2008hk,Tegmark:2003ud,lcdmfits}
that the simplest cosmological model consistent with the observed
accelerating expansion is the $\Lambda$CDM model \cite{lcdmrev}
(based on a cosmological constant) even though models based on
dynamical dark energy \cite{dde} or modified gravity
\cite{Boisseau:2000pr} remain viable alternatives.

Several new SnIa datasets have emerged during the last 3-4 years
aiming at mapping in detail the accelerating expansion in the
redshift range $z\in [0,2]$. The main such datasets, analyzed in
the present study, are shown in detail in Table \ref{tabsnsets} of
the next section and include the compilations: Constitution
\cite{Hicken:2009dk}, Union \cite{Kowalski:2008ez}, ESSENCE
\cite{Davis:2007na}, SNLS 1st year (SNLS1) \cite{Astier:2005qq},
Gold06 \cite{Riess:2006fw} and SDSS-II \cite{Kessler:2009ys}. As
shown in Table \ref{tabsnsets} most of these compilations include
subsets obtained with different instruments. Even though efforts
have been made in most cases
to reanalyze the SnIa light curves and reject outliers in order to
smooth out potential systematics due to inhomogeneities of the
samples, systematics remain as a source of uncertainties \cite{Nesseris:2006ey,Wei:2009ry}.

The above SnIa standard candle data (luminous sources of known
intrinsic luminosity) are geometric probes used to measure the
luminosity distance $d_L(z)$ which, assuming flatness, is
connected to the Hubble expansion rate $H(z)$ as \be d_L (z)= c
(1+z) \int_0^z dz'\frac{1}{H(z')}\,, \label{dl1} \ee where $c$ is
the velocity of light. Alternative geometric probes are standard
rulers (objects of known comoving size). These may be used to
measure the angular diameter distance $d_A(z)$ which, in a flat
universe, is related to $H(z)$ as \be d_A (z)= \frac{c}{1+z}
\int_0^z dz'\frac{1}{H(z')} \label{da1}. \ee The most useful
standard ruler in cosmology is the last scattering sound horizon
($z\simeq 1090$), the scale of which can be measured either
directly through the CMB temperature power spectrum
\cite{Spergel:2006hy,Hu:2000ti,Komatsu:2008hk} or indirectly
through Baryon Acoustic Oscillations (BAO) on the matter power
spectrum at low redshifts
\cite{Eisenstein:2005su,Percival:2007yw,Percival:2009xn}. These
data lead to constraints on $H(z)$ that are independent from those
of standard candles. Thus the mutual consistency of the standard
candle and standard ruler constraints can be used as a quality
test for both classes of data. This is one of the consistency
tests implemented on SnIa datasets in the present study.

All of the above mentioned geometric probes aim at mapping the
expansion rate $H(z)$ as a function of the redshift $z$. The
determination of the Hubble parameter H(z) is equivalent to
identifying the function $w(z)$ defined as \ba
w(z)\,&=&\,-1\,+\frac{1}{3}(1+z)\cdot
\\&&\frac{d\ln (H(z)^2/H_0^2-\Omega_{\rm 0m}
(1+z)^3-\Omega_{0r} (1+z)^4)}{d z}\nn{\label{wzh1}} \ea where the
term in the logarithm accounts for all terms in the Friedmann
equation not related to matter (present normalized density $\omm$)
and radiation (present normalized density $\Omega_{0r}$). If the
origin of the accelerating expansion is dark energy then $w(z)$
may be identified with the dark energy equation of state parameter
$w(z)=\frac{p_X}{\rho_X}$. The cosmological constant ($w(z)=-1$)
is the simplest dark energy model and corresponds to a constant
dark energy density ($\Lambda$CDM model). The vast majority of
presently available cosmological data are consistent with
$\Lambda$CDM \cite{lcdmfits} at the $95.4\%$ level (see however
\cite{Perivolaropoulos:2008ud} and references therein for some
puzzling exceptions).

In view of the increasing number of SnIa dataset compilations a
need has emerged for ranking these compilations with respect to
their Figure of Merit \cite{Albrecht:2006um, Linder:2006xb} (FoM),
defined as the inverse area of the $95\%$ confidence region
\footnote{Note that we use the $2\sigma$ contour in parameter
space instead of the $95\%$ confidence region used by the Dark
Energy Task Force.}, their degree of consistency with
$\Lambda$CDM, with standard candles and with each other. These
consistency rankings require the derivation of suitable statistics
designed to achieve them in an efficient manner. The goal of the
present study is to provide such statistics and apply them in
order to rank the SnIa compilations of Table \ref{tabsnsets} of
section 2 according to
\begin{itemize} \item their FoM in the context of the Chevalier-Polarski-Linder (CPL) \cite{Chevallier:2000qy},
\cite{Linder:2002et} parametrization of dynamical dark energy \be
w=w_0+w_1(1-a)=w_0+w_1\frac{z}{1+z} \label{cpl} \ee \item their degree of consistency with
$\Lambda$CDM \item their degree of consistency with standard ruler
constraints of Ref. \cite{Komatsu:2008hk,Percival:2007yw} thus testing the
quality of the SnIa data (assuming that the distance duality
relation ${d_L(z)}={d_A(z) (1+z)^2}$ is applicable) \item their
degree of consistency with each other. \end{itemize}

In the context of ranking the SnIa compilations with respect to
their consistency with $\Lambda$CDM we make use of the CPL
parametrization and assuming flatness, we identify the
``distance'' in units of $\sigma$ ($\sigma$-distance $d_\sigma
(\omm)$) of the ``reference'' parameter space point
$(w_0,w_1)=(-1,0)$ corresponding to $\Lambda$CDM from the best fit
point $(w_0,w_1)$ of each SnIa dataset and for several priors of
$\Omega_{\rm 0m}$. Similarly, in order to rank the SnIa
compilations with respect to their consistency with CMB-BAO
standard rulers we follow the above method but we replace the
$\Lambda$CDM ``reference'' point by the best fit $(w_0,w_1)^{SR}$
parameter values obtained in the context of standard ruler
(CMB+BAO) data for each $\omm$ prior.

In addition to the $\sigma$-distance we also use the Binned
Normalized Differences (BND) statistic of Ref.
\cite{Perivolaropoulos:2008yc} to rank the datasets according to
their consistency with $\Lambda$CDM. Finally, in the context of measuring
the internal consistency of a set of n compilations we consider the
mean $\sigma$-distance \be {\bar d}_{\sigma} (\omm; w_0,
w_1)\equiv \frac{1}{n}\sum_{i=1}^n d_{\sigma i} (\omm; w_0,
w_1)\,, \label{mtd} \ee where $d_{\sigma i} (\omm; w_0, w_1)$ is
the $\sigma$-distance from the reference parameter point $(w_0,
w_1)$ to the best fit point of the $i^{th}$ compilation (see
Fig.~\ref{figg}). Minimization of ${\bar d}_{\sigma} (\omm; w_0,
w_1)$ at fixed $\omm$ leads to a minimum mean $\sigma$-distance
and to the parameters $({\bar w}_0, {\bar w}_1)$ of maximum
consistency for the particular set of compilations. Smaller ${\bar
d}_{\sigma} (\omm; {\bar w}_0, {\bar w}_1)$ implies a better
internal consistency for the set of compilations. Minimizing ${\bar
d}_{\sigma} (\omm; {\bar w}_0, {\bar w}_1)$ with respect to $\omm$
we can also find the value of $\omm$ that maximizes the
consistency among the SnIa data compilations.

The above statistics are applied on the SnIa data of Table
\ref{tabsnsets} in the following sections. In section 2 we
summarize the likelihood calculations needed to evaluate the
$\sigma$-distances described above and briefly discuss the BND
statistic described in detail in Ref.
\cite{Perivolaropoulos:2008yc}. We also discuss the main features
of the considered SnIa datasets and rank them according to their
FoM. In section 3 we apply the $\sigma$-distance statistic and the
BND statistic to rank the data of Table \ref{tabsnsets} according
to their consistency with $\Lambda$CDM and with the CMB-BAO best
fit. In section 4 we apply the mean $\sigma$-distance minimization
to find the internal consistency of various sets of compilations
obtained from Table \ref{tabsnsets}. Finally in section 5 we
conclude, summarize and discuss future prospects of the present
study.

\section{Likelihood Analysis}
Our likelihood analysis method is described in detail in Refs
\cite{Nesseris:2006er} and \cite{Lazkoz:2007cc}. Here we only
review some of the basic steps for completeness. We assume a CPL
\cite{Chevallier:2000qy}, \cite{Linder:2002et} parametrization for
$w(z)$ (equation (\ref{cpl})) and apply the maximum likelihood
method separately for standard rulers (CMB+BAO) and standard
candles (SnIa) assuming flatness. The corresponding late time form
of $H(z)$ for the CPL parametrization is \ba H^2 (z)&=&H_0^2 [
\Omega_{\rm 0m} (1+z)^3 + \nn
\\ &+& (1-\Omega_{\rm 0m})(1+z)^{3(1+w_0+w_1)}e^{\frac{-3w_1
z}{(1+z)}}]\,. \label{hcpl} \ea

In the context of constraints from standard rulers we use the
datapoints $(R,l_a,100\obh)$ of Ref. \cite{Komatsu:2008hk} (WMAP5)
where $R$, $l_a$ are two shift parameters \cite{Lazkoz:2007cc}.

For a flat prior, the 5-year WMAP data (WMAP5) measured best fit
values are \cite{Komatsu:2008hk}
\begin{eqnarray}
\hspace{-.5cm}\bf{{\bar V}_{CMB}} &=& \left(\begin{array}{c}
{\bar R} \\
{\bar l_a}\\
{100\obh}\end{array}
  \right)=
  \left(\begin{array}{c}
1.710\pm 0.019 \\
302.10 \pm 0.86\\
2.2765 \pm 0.0596 \end{array}
  \right)
\label{cmbdat} \end{eqnarray} The corresponding covariance matrix
is \cite{Komatsu:2008hk}

\begin{eqnarray}
 {\bf C_{CMB}}=\left(
\begin{array}{ccc}
0.000367364& 0.00181498& -0.000201759\\
0.00181498& 0.731444& -0.0315874\\
-0.000201759& -0.0315874& 0.00355323
\end{array}
\right)\nn \\
\end{eqnarray}

We thus define
\begin{eqnarray}
\bf{X_{CMB}} &=& \left(\begin{array}{c}
R - 1.710 \\
l_a-302.10\\
100\Omega_bh^2-2.2765\end{array}
  \right)
\end{eqnarray}
and construct the contribution of CMB to the $\chi^2$ as \be
\chi^2_{CMB}=\bf{X_{CMB}}^{T}{\bf C_{CMB}}^{-1}\bf{X_{CMB}} \ee
with the inverse covariance matrix
\begin{eqnarray}
\hspace{-.5cm} {\bf C_{CMB}}^{-1}=\left(
\begin{array}{ccc}
2809.73& -0.133381& 158.356\\
-0.133381& 2.21908& 19.7195\\
158.356& 19.7195& 465.728
\end{array}
\right)
\end{eqnarray}
Notice that $\chi^2_{CMB}$ depends on the four parameters
($\omms$, $\ob$, $w_0$ and $w_1$). In what follows we use the
priors $h=0.705$, $\ob=2.2765/100 h^2$ \cite{Komatsu:2008hk}.

In the case of BAO we also apply the maximum likelihood method
\cite{Lazkoz:2007cc} using the datapoints of  Ref.
\cite{Percival:2007yw} (SDSS5). For comparison, we have also
considered the more recent data of Ref. \cite{Percival:2009xn}
(SDSS7) and have found minor differences in the results (slightly
reduced consistency with \lcdm in the context of the CPL
parametrization but no change in the ranking sequences). In some
cases we show the results of both sets of datapoints
(Figs.~3,~4a,~9a).

\vspace{0pt}
\begin{table*}[t!]
\begin{center}
\caption{The datasets used in the present analysis. See respective
references for details on the sources of the SnIa data points.
\label{tabsnsets}}
\begin{tabular}{cccccc}
\hline
\hline\\
\vspace{6pt}\textbf{Dataset} \hspace{7pt}& \textbf{Date Released}\hspace{7pt}& \textbf{Redshift Range} \hspace{7pt} & \textbf{\# of SnIa}\hspace{7pt} & \textbf{Filtered subsets included}\hspace{7pt}                                                       & \textbf{Refs }  \\
\vspace{6pt} SNLS1       & 2005 \hspace{7pt} & $0.015\leq z \leq 1.01$ \hspace{7pt} & 115  \hspace{7pt} & SNLS \cite{Astier:2005qq}, LR \cite{lr}                                                     & \cite{Astier:2005qq} \\
\vspace{6pt} Gold06      & 2006 \hspace{7pt} & $0.024\leq z \leq 1.76$ \hspace{7pt} & 182  \hspace{7pt} & SNLS1 \cite{Astier:2005qq}, HST \cite{Riess:2006fw}, SCP \cite{scp}, HZSST \cite{hzsst}  & \cite{Riess:2006fw} \\
\vspace{6pt} ESSENCE     & 2007 \hspace{7pt} & $0.016\leq z \leq 1.76$ \hspace{7pt} & 192  \hspace{7pt} & SNLS1 \cite{Astier:2005qq}, HST \cite{Riess:2006fw}, ESSENCE\cite{WoodVasey:2007jb},\cite{Davis:2007na}         & \cite{WoodVasey:2007jb},\cite{Davis:2007na} \\
\vspace{6pt} Union       & 2008 \hspace{7pt} & $0.015\leq z \leq 1.55$ \hspace{7pt} & 307  \hspace{7pt} & Gold06 \cite{Riess:2006fw}, ESSENCE\cite{WoodVasey:2007jb},\cite{Davis:2007na}                                    & \cite{Kowalski:2008ez} \\
\vspace{6pt} Constitution& 2009 \hspace{7pt} & $0.015\leq z \leq 1.55$ \hspace{7pt} & 397  \hspace{7pt} & Union \cite{Kowalski:2008ez}, CfA3\cite{Hicken:2009dk}                                     & \cite{Hicken:2009dk} \\
\vspace{6pt} SDSS     & 2009 \hspace{7pt} & $0.022\leq z \leq 1.55$ \hspace{7pt} & 288  \hspace{7pt} & \begin{tabular}{c} Nearby \cite{Jha:2007ai}, SDSS-II \cite{Kessler:2009ys}, ESSENCE \cite{WoodVasey:2007jb},\\SNLS \cite{Astier:2005qq}, HST \cite{Riess:2006fw}\end{tabular}  & \cite{Kessler:2009ys} \\
\hline \hline
\end{tabular}
\end{center}
\end{table*}

Following Ref. \cite{Lazkoz:2007cc} we find the contribution of BAO to $\chi^2$ as \be
\chi^2_{BAO}=\bf{X_{BAO}}^{T}{\bf C_{BAO}}^{-1}\bf{X_{BAO}}\,. \ee

The analysis of SnIa standard candles is also based on the method
described in Ref. \cite{Lazkoz:2007cc}. Two of the earliest and
reliable datasets are the Gold06 dataset \cite{Riess:2006fw} and
the Supernova Legacy Survey (SNLS) \cite{Astier:2005qq} dataset.
The Gold06 dataset compiled by Riess et. al. is a set of supernova
data from various sources analyzed in a consistent and robust
manner with reduced calibration errors arising from systematics.
It contains 152 points from previously published data plus 30
points with $z>1$ discovered by the HST\cite{Riess:2006fw}. Even
though the Gold06 data are of high quality, they are plagued by
non-uniformity (they are a collection of data obtained from
various instruments) and they include a few outliers. Filtered
versions of the Gold06 dataset attempting to deal with these
problems have been included in subsequent compilations.

\begin{figure*}[!t]
\rotatebox{0}{\hspace{0cm}\resizebox{0.45\textwidth}{!}{\includegraphics{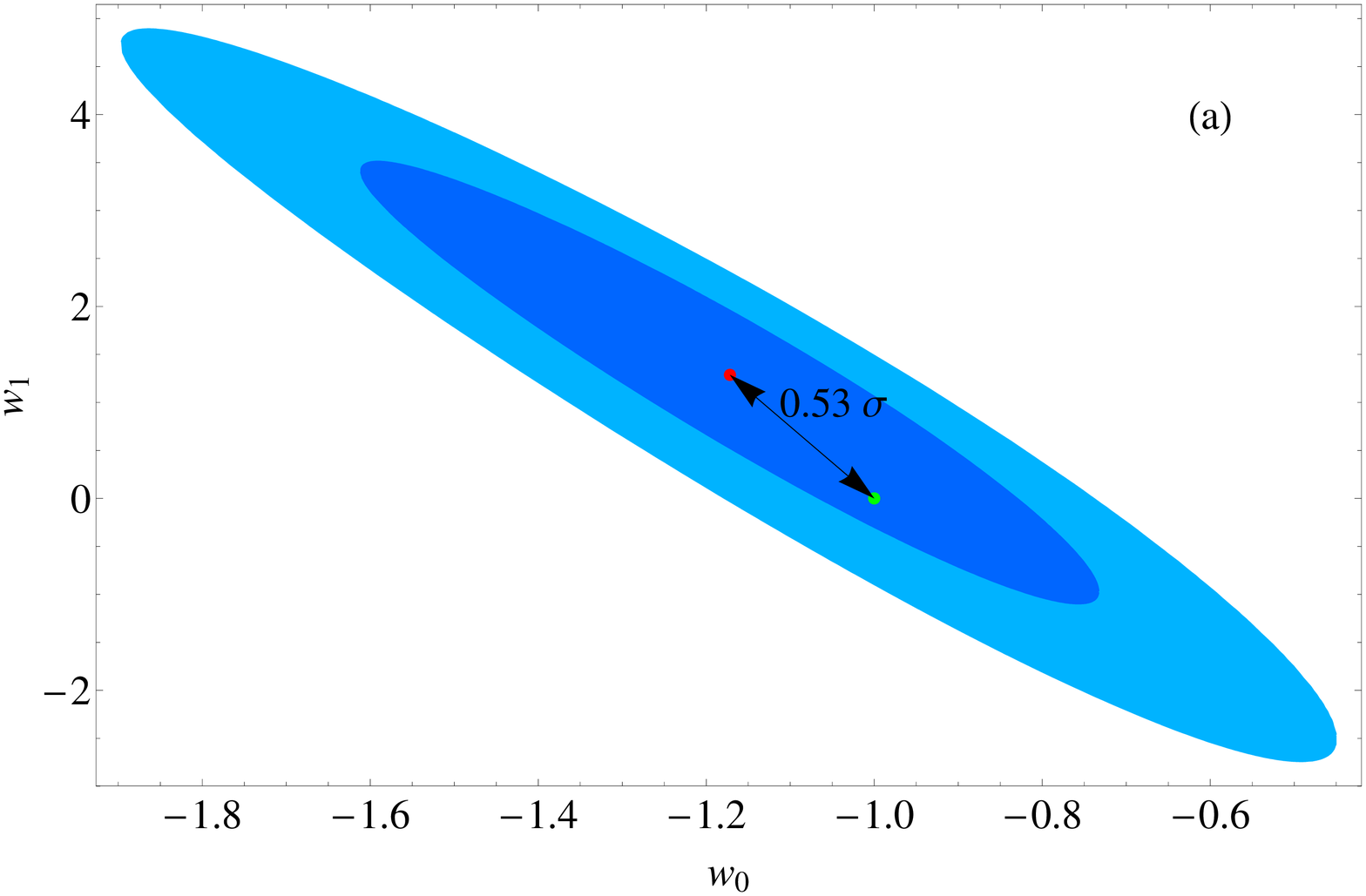}}}
\rotatebox{0}{\hspace{1cm}\resizebox{0.45\textwidth}{!}{\includegraphics{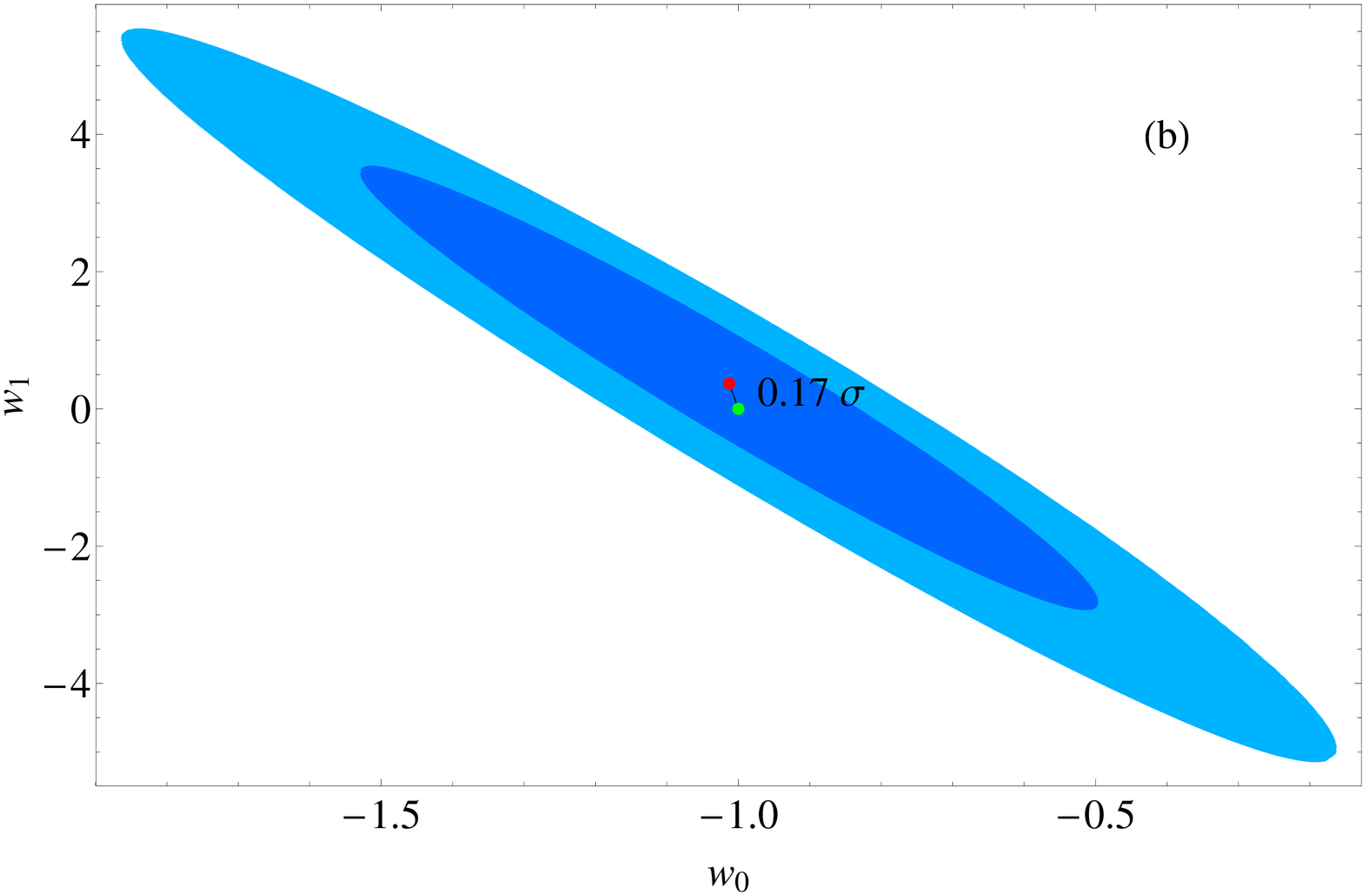}}}
\vspace{0pt}\caption{The $68.3\% (1\sigma)-95.4\%$ ($2\sigma$) $\chi^2$ confidence
contours in the $w_{0}-w_{1}$ plane based on parametrization
(\ref{hcpl}) for the ESSENCE (left) and SNLS1 datasets (right) for
$\omm=0.24$. The arrows indicate the $\sigma$-distance of $\Lambda$CDM
(green points: $(w_0,w_1)=(-1,0)$) to the best fit points (red
points). }\label{fig1}
\end{figure*}

The SNLS is a 5-year survey of SnIa with $z<1$. The SNLS has
adopted a more efficient SnIa search strategy involving a
``rolling search'' mode where a given field is observed every
third or fourth night using a single imaging instrument, thus
reducing photometric systematic uncertainties. The published first
year SNLS dataset (SNLS1) constitutes of 44 previously published
nearby SnIa with $0.015<z<0.125$ plus 73 distant SnIa ($0.15<z<1$)
discovered by SNLS, two of which are outliers and are not used in
the analysis. At this point it should be mentioned that both SNLS
and Gold06 are reliable datasets, however, the non-uniformity of
the Gold06 makes it less reliable compared to SNLS which is
significantly more uniform.

We will also use the ESSENCE SnIa dataset of Davis et. al.
\cite{Davis:2007na} which constitutes of four subsets: ESSENCE
\cite{essence,WoodVasey:2007jb} (60 points), SNLS1
\cite{Astier:2005qq} (57 points), nearby \cite{Riess:2004nr} (45
points) and HST \cite{Riess:2006fw} (30 points) and the Union SnIa
dataset of Kowalski et. al. \cite{Kowalski:2008ez} which
constitutes of 414 SnIa, reduced to 307 points after various
selection cuts were applied in order to create a homogeneous and
high-signal-to-noise dataset. Finally, we will also use the
Constitution SnIa dataset of Hicken et. al. \cite{Hicken:2009dk}
which constitutes in total of 397 SnIa out of which ~100 come from
the new low-$z$ CfA3 sample and the rest from the Union
\cite{Kowalski:2008ez} dataset. The inclusion of the new low-$z$
sample in the Constitution dataset is a major improvement over
previous datasets because the previous sample of nearby SnIa was
relatively small and based on early investigations, leading to
significant systematic uncertainties. In our analysis we use the
Constitution datasets analyzed with both SALT and MLCS17
fitters~\cite{Hicken:2009dk}.

Finally, we also consider the recently released first year data of
the SDSS-II \cite{Kessler:2009ys} aiming to alleviate the lack of
data at intermediate redshift. In our analysis we consider the
largest combined sample of SnIa considered in
\cite{Kessler:2009ys}. This encompasses 103 SnIa from the SDSS-II,
33 nearby SnIa \cite{Jha:2007ai}, 56 points from ESSENCE
\cite{WoodVasey:2007jb}, 62 from SNLS \cite{Astier:2005qq} and 34
from HST \cite{Riess:2006fw} making up a total of 288 SnIa. Here,
we consider the SDSS-II dataset analyzed with both the SALT2
light-curve fitter and the MLCS2k2 fitter\cite{fitters} and we
compare the corresponding consistencies with \lcdm. We have made
simple tests to ensure that our analysis agrees with the one from
\cite{Kessler:2009ys}, but in the case of the SDSS-II dataset
analyzed with the SALT2 fitter, there is some ambiguity as the
authors of Ref.~\cite{Kessler:2009ys} have not released the
necessary covariance matrix that accounts for the correlations
between the shape luminosity parameter ($x_1$), the color
parameter ($c$) and the overall flux normalization ($x_0$).

In Table \ref{tabsnsets} we give some details about the SnIa
datasets used in this analysis, such as the redshift range or the
subsets of each set.

The SnIa observations use proper light curve fitters\cite{fitters}
to provide the apparent magnitude $m(z)$ of the supernovae at peak
brightness. The resulting apparent magnitude $m(z)$ is related to
the dimensionless luminosity distance $D_L(z)$ through \be
m_{th}(z)={\bar M} (M,H_0) + 5 log_{10} (D_L (z))\,, \label{mdl}
\ee where we have used the notation of Ref. \cite{Lazkoz:2007cc}.
The theoretical model parameters are determined by minimizing the
quantity \be \chi^2_{SnIa} (\om,w_0,w_1)= \sum_{i=1}^N
\frac{(\mu_{obs}(z_i) - \mu_{th}(z_i))^2}{\sigma_{\mu \; i}^2 }
\label{chi2} \ee where $N$ is the number of SnIa of the dataset
and $\sigma_{\mu \; i}^2$ are the errors due to flux
uncertainties, intrinsic dispersion of SnIa absolute magnitude and
peculiar velocity dispersion. These errors are assumed to be
Gaussian and uncorrelated. The theoretical distance modulus is
defined as \be \mu_{th}(z_i)\equiv m_{th}(z_i) - M =5 log_{10}
(D_L (z)) +\mu_0\,, \label{mth} \ee where \be \mu_0= 42.38 - 5
log_{10}h\,, \label{mu0}\ee  The steps we followed for the usual
minimization of (\ref{chi2}) in terms of its parameters are
described in detail in Refs.
\cite{Nesseris:2004wj,Nesseris:2005ur,Nesseris:2006er}.

\vspace{0pt}
\begin{table}[!t]
\begin{center}
\caption{The Figure of Merit (FoM) for the datasets of Table
\ref{tabsnsets} for $\omm=0.28$ using the CPL parametrization. For
comparison we also show the corresponding FoM obtained with
standard ruler data. \label{tabsfom}}
\begin{tabular}{ccccc}
\hline
\hline\\
\vspace{6pt}\textbf{Dataset} \hspace{7pt}& \textbf{\# of SnIa}\hspace{7pt} & \textbf{Figure of Merit}\hspace{7pt} \\
\vspace{6pt} SNLS1                    & 115                \hspace{7pt} &  0.208                               \\
\vspace{6pt} Gold06                      & 182                \hspace{7pt} &  0.367                               \\
\vspace{6pt} ESSENCE                     & 192                \hspace{7pt} &  0.245                               \\
\vspace{6pt} SDSS-II (SALT2)                  & 288                \hspace{7pt} &  0.366                               \\
\vspace{6pt} SDSS-II (MLCS2k2)                  & 288                \hspace{7pt} &  0.553                               \\
\vspace{6pt} Union                       & 307                \hspace{7pt} &  0.512                               \\
\vspace{6pt} Constitution (SALT2)               & 397                 \hspace{7pt} &  0.708                               \\
\hline \\
\vspace{6pt} CMB+SDSS5                & -                \hspace{7pt} & 2.028                               \\
\vspace{6pt} CMB+SDSS7                & -                \hspace{7pt} & 2.541                              \\
\hline \hline
\end{tabular}
\end{center}
\end{table}

In order to study the consistency of the various SnIa datasets
with the cosmological constant and the standard rulers we consider
the distance in units of $\sigma$ ($\sigma$-distance $d_\sigma$)
of the best fit point to a model with parameters $(w_0,w_1)$,
where this reference point can be either $\Lambda$CDM or some
other reference point, (see Fig.~\ref{fig1}). The
$\sigma$-distance can be found by converting $\Delta
\chi^2=\chi^2_{(w_0,w_1)}-\chi^2_{min}$ to $d_\sigma$, i.e.
solving \cite{press92} \be 1-\Gamma(1,\Delta
\chi^2/2)/\Gamma(1)={\rm Erf}(d_\sigma/\sqrt{2}) \label{sigmas}\ee
for $d_\sigma$ ($\sigma$-distance), where $\Delta \chi^2$ is the
$\chi^2$ difference between the best-fit and the reference point
($w_0,w_1)$ (eg $\Lambda$CDM) and ${\rm Erf}()$ is the error
function. The right hand side of Eq.~(\ref{sigmas}) comes from
integrating $\int_{-n \sigma}^{~n \sigma}\frac{1}{\sigma \sqrt{2
\pi}}~e^{-\frac{x^2}{2 \sigma^2}}dx$, where $n$ is the desired
number of $\sigma$s, while the left hand side corresponds to the
Cumulative Distribution Function (CDF) of a $\chi^2$
distribution\cite{press92} with two degrees of freedom. Note that
Eq.~(\ref{sigmas}) is only valid for the two parameters
$(w_0,w_1)$ and should be generalized accordingly for more
parameters \cite{press92}. In the special case of $n=1$ or $n=2$
we obtain the well known results $\Delta \chi_{1\sigma}=2.30$ and
$\Delta \chi_{2\sigma}=6.18$ valid for two parameter
parametrizations \cite{press92}.

Even though the $\sigma$-distance is not a commonly used statistic it
is quite useful because it can directly give information about
probability of a given region in parameter space. The integer
values of sigma distance ($1\sigma$ and $2\sigma$) are commonly
used to draw the corresponding contours in parameter space. We
have extended this statistic to non-integer values in order to
find the specific contours that go through particular reference
points of parameter space and thus estimate quantitatively the
consistency of these points. The advantage of using the
$\sigma$-distance instead of $\Delta\chi^2$ is the fact that the
$\sigma$-distance takes into account the number of parameters of
the parameterizations and can therefore be directly translated
into probability for each point in parameter space. This is not
possible for $\Delta\chi^2$ because it does not include
information about the number of parameters of the
parameterizations considered.

\begin{figure}[!t]
\vspace{0cm}\includegraphics[width=0.48\textwidth]{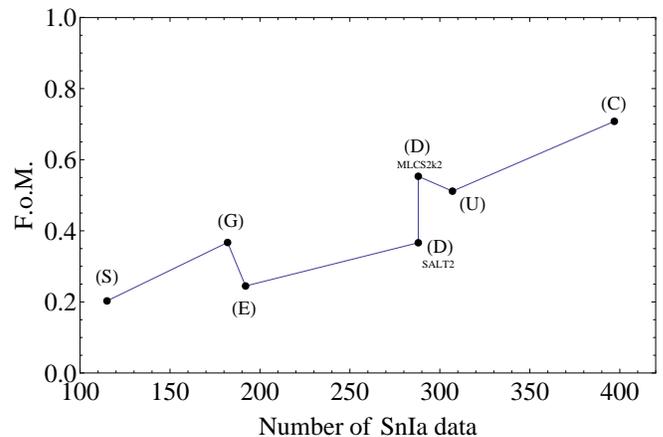}
\caption{The Figure of Merit (FoM) in terms of the number of the
SnIa data for $\omm=0.28$ using the CPL
parametrization.\label{fig1c} }
\end{figure}

The Figure of Merit (FoM) is a useful measure of the effectiveness
of a set of data in constraining cosmological parameters. In the
case of two parameters (as for the CPL parametrization) it is
defined as the reciprocal area of the $95.4\%$ contour, in
parameter space $(w_0,w_1)$ \cite{Albrecht:2006um,Linder:2006xb}.
Clearly, the larger the FoM the more effective the dataset in
constraining the parameters $(w_0,w_1)$. In Table \ref{tabsfom} we
show the FoM for each dataset of Table 1 for a prior of
$\omm=0.28$. In Fig.~\ref{fig1c} we show the FoM in terms of the
number of the SnIa data for the same datasets. Clearly, the FoM is
an increasing function of the number of SnIa in the datasets. An
exception to this rule is the ESSENCE dataset which has a slightly
smaller FoM compared to the Gold06 dataset even though it has a
larger number of SnIa.

We should stress that the FoM does not depend only on the total
number of SnIa of the dataset but mainly on the number of SnIa at
low and high redshifts. This sensitivity on the distribution in
redshift space is most probably the origin of the ESSENCE glitch
in the FoM plot of Fig.~\ref{fig1c}. Notice that the redshift
space distribution of the ESSENCE data includes more data at
intermediate redshifts than the Gold06 dataset while the number of
SnIa in the ESSENCE dataset is similar to that of the Gold06
dataset. Finally, for comparison, in the same table we show the
FoM corresponding to standard ruler data (WMAP5+SDSS5 and
WMAP5+SDSS7). Clearly, the FoM of standard rulers is about a
factor of 3 higher compared to the highest FoM of SnIa
corresponding to the Constitution dataset.

In addition to using $d_\sigma$ to rank SnIa datasets, we consider
the BND statistic \cite{Perivolaropoulos:2008yc} which is designed
to pick up systematic brightness trends of SnIa datapoints with
respect to a best fit cosmological model at high redshifts. It is
based on binning (considering the average of) the normalized
differences between the SnIa distance moduli and the corresponding
best fit values in the context of a specific cosmological model
(e.g. $\Lambda$CDM). These differences are normalized by the
standard errors of the observed distance moduli (BND). As in Ref.
\cite{Perivolaropoulos:2008yc} we will focus on the highest
redshift bin and extend its size towards lower redshifts until the
BND changes sign (crosses 0) at a redshift $z_c$ (bin size $N_c$).
The bin size $N_c$ of this crossing (the statistical variable) is
then compared with the corresponding crossing bin size $N_{mc}$
for Monte Carlo data realizations based on the best fit model.

\begin{figure}[!t]
\vspace{-0.2cm}\includegraphics[width=0.55\textwidth]{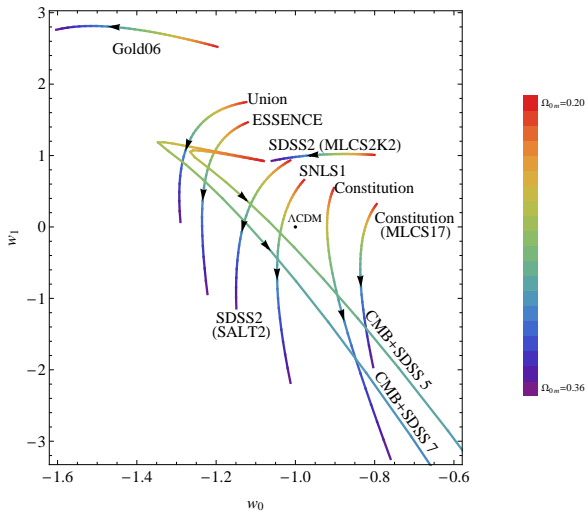}
\caption{Trajectories of the best fit points $(w_0,w_1)$ obtained
for each of the datasets of Table \ref{tabsnsets} and also for the
standard ruler CMB-BAO (WMAP5+SDSS5 and WMAP5+SDSS7) data as
$\omm$ varies in the range $\omm \in [0.2,0.36]$. The arrows in
the best fit lines indicate the direction of growing $\omm$. Note
that for the SDSS5 data the standard ruler best fit parameters
stretch out to $(w_0,w_1)\simeq(2,-30)$ for $\omm\simeq 0.36$,
whereas for the SDSS7 data
$(w_0,w_1)\simeq(0.90,-20)$.}\label{fig2}
\end{figure}

In Ref. \cite{Perivolaropoulos:2008yc} it was found that the
crossing bin size $N_c$ obtained from the Union and Gold06 data
with respect to the best fit $\Lambda$CDM model is anomalously
large compared to $N_{mc}$ of the corresponding Monte Carlo
datasets obtained from the best fit $\Lambda$CDM in each case. In
the next section we will extend this analysis to all the datasets
of Table \ref{tabsnsets} and use the results to rank these
datasets according to their consistency with $\Lambda$CDM.

\section{Consistency of Datasets with $\Lambda$CDM and with Standard Rulers}
It is straightforward to apply the likelihood methods described in
the previous section to find the trajectory of the best fit point
$(w_0,w_1)$ in parameter space as $\omm$ varies in the range $\omm
\in [0.2,0.36]$. These trajectories obtained for each of the
datasets of Table \ref{tabsnsets} and also for the standard ruler
CMB-BAO (WMAP5+SDSS5, WMAP5+SDSS7) data are shown in
Fig.~\ref{fig2}. These trajectories can not be used to directly
rank the datasets according to their consistency with any given
reference point in parameter space (e.g. $\Lambda$CDM) because
they contain no information about the $68.3\%$ contours (the plot
would become confusing if we attempted to include such contours).
However, they provide useful hints for the trend of the best fit
parameters as $\omm$ varies. For example, such a trend is the
increase of the best fit value of the slope $w_1$ as the prior of
$\omm$ decreases towards the value $0.2$ or that the best fit
value of $w_0$ remains less than $-1$ for all datasets except of
the Constitution compilation.
\begin{figure*}[!t]
\centering
\begin{center}
$\begin{array}{@{\hspace{-0.10in}}c@{\hspace{0.0in}}c}
\multicolumn{1}{l}{\mbox{}} & \multicolumn{1}{l}{\mbox{}} \\
[-0.2in] \epsfxsize=3.3in \epsffile{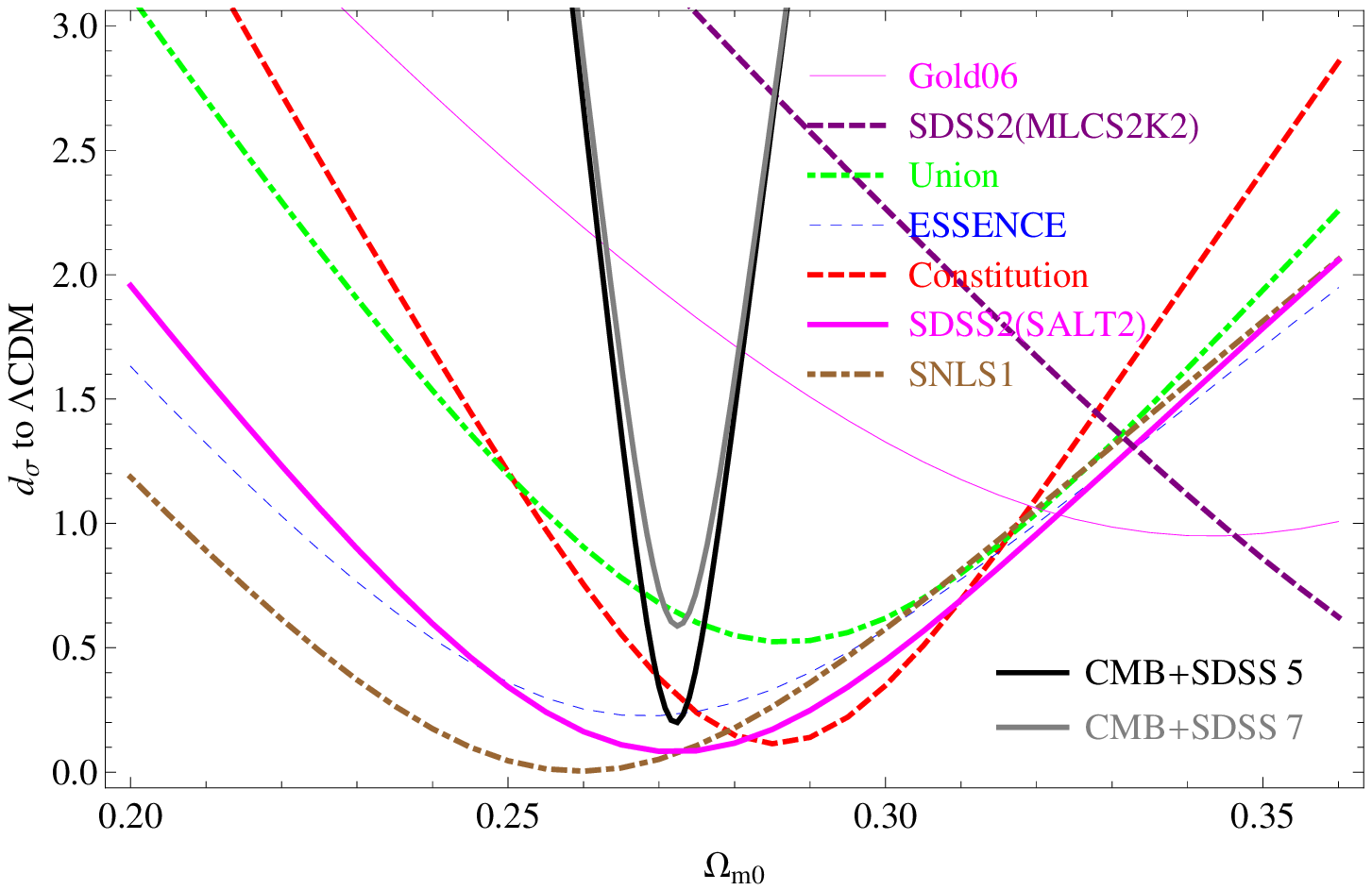} & \epsfxsize=3.3in
\epsffile{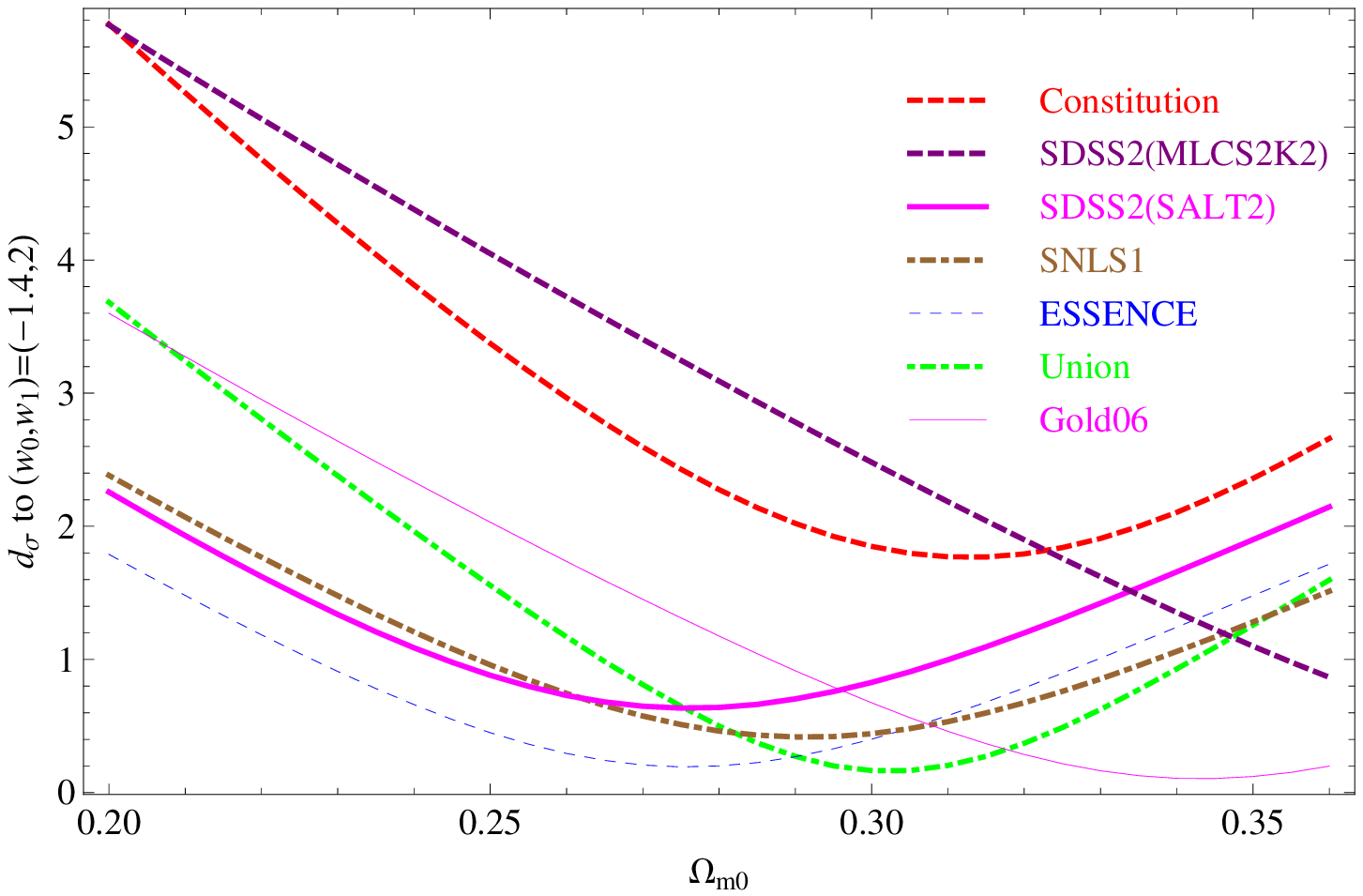} \\
\end{array}$
\end{center}
\vspace{0.0cm} \caption{\small a: $\sigma$-distances $d_{\sigma
i}(\omm;-1,0)$ off the $\Lambda$CDM point (reference point) from
the best fit of each dataset in Table \ref{tabsnsets}. Notice that
they are minimized at similar values of $\omm$. The
$\sigma$-distance $d_\sigma(\omm;-1,0)$ between the standard ruler
best fits and $\Lambda$CDM are also shown as a function of $\omm$
(black and grey solid lines).  b: Similar to (a) for the dynamical
dark energy reference point $(w_0, w_1)=(-1.4,2)$. Notice that the
$\sigma$-distances $d_{\sigma i}(\omm;-1.4,2)$ are minimized at
more widely separated values of $\omm$.} \label{fig3}
\end{figure*}

The ranking sequence of the datasets of Table \ref{tabsnsets} with
respect to any reference point in parameter space can be studied
quantitatively using the $\sigma$-distance statistic discussed
above. In order to test the sensitivity of the ranking sequence of
datasets with respect to the choice of consistency reference point
$(w_0, w_1)$ we consider two such reference points: $(w_0,
w_1)=(-1,0)$ ($\Lambda$CDM) and $(w_0, w_1)=(-1.4,2)$ which
corresponds to dynamical dark energy with a $w(z)$ that crosses
the line $w=-1$. It should be noted that there is nothing special
about the parameter point (-1.4,2). We have selected it as a
representative of a wide region in parameter space (upper left
from $\Lambda$CDM) which corresponds to dynamical dark energy
crossing the phantom divide line $w=-1$. Any other point in the
same parameter region would lead to similar results and the same
ranking of datasets. This particular parameter region is
interesting because it is spanned by the best fit trajectories and
it also mildly favored by the Gold06 dataset (see Fig. 3).

The resulting $d_\sigma(\omm)$ for each dataset of Table
\ref{tabsnsets} are shown in Figs.~\ref{fig3}a and \ref{fig3}b in
the range $\omm \in [0.2, 0.36]$.
\begin{table}[!t]
\begin{center}
{\renewcommand{\arraystretch}{1.3}
\begin{tabular}{|@{\hspace{0.25cm}}c@{\hspace{0.25cm}}|@{\hspace{0.25cm}}
c@{\hspace{0.25cm}}c@{\hspace{0.25cm}}c@{\hspace{0.25cm}}c@{\hspace{0.25cm}}
|}\hline Dataset & $d_\sigma^{min}$ & $\Omega_{0m}^{min}$ & $w_0$
& $w_1$\\\hline
SNLS1 & 0.004 & 0.260 & -1.03 & 0.16 \\
SDSS-II (SALT2) & 0.084 & 0.270 & -1.09 & 0.51 \\
Constitution & 0.114 & 0.285 & -0.91 & -0.54 \\
ESSENCE & 0.227 & 0.270 & -1.20 & 1.04 \\
Union & 0.525 & 0.285 & -1.25 & 1.40 \\
SDSS-II (MLCS2K2) & 0.623 & 0.360 & -1.06 & 0.93 \\
Gold06 & 0.950 & 0.345 & -1.56 & 2.80 \\\hline\hline CMB+BAO
(SDSS5) & 0.200 & 0.272 & -1.15 & 0.51\\ CMB+BAO (SDSS7) & 0.588 &
0.272 & -1.30 & 0.97\\\hline
\end{tabular}}
\end{center}
\caption{Minimum $\sigma$-distance
$d_\sigma^{min}(\omm^{min};-1,0)$ from the best fit point for each
of the datasets to the $\Lambda$CDM point. Also listed are the
corresponding values of $\omm$, and the best fit parameters
$(w_0,w_1)$ (see also Fig.~\ref{fig3}a). The SDSS-II (MLCS2K2)
data showed no minimum of $d_\sigma$ with respect to $\omm$ in the
range $\omm \in [0.2,0.36]$. We thus have simply displayed the
lowest value of $d_\sigma$ in the corresponding range of
$\omm$.}\label{tabsep1}
\end{table}
Clearly, there are values of $\omm$ that minimize the
$\sigma$-distance $d_\sigma(\omm)$ between the best fit of each
dataset and the reference point. These values of $\omm$ maximize
the consistency of the datasets with the given reference point in
this range of $\omm$. The minima $\sigma$-distances
$d_\sigma(\omm)$ for each dataset, corresponding to maximum
consistency with $\Lambda$CDM along with the corresponding values
of $\omm$ are shown (properly ranked) in Table \ref{tabsep1}. The
corresponding results for the reference point $(w_0,
w_1)=(-1.4,2)$ are shown in Table \ref{tabsep2}.

\begin{table}[!t]
\begin{center}
{\renewcommand{\arraystretch}{1.3}
\begin{tabular}{|@{\hspace{0.25cm}}c@{\hspace{0.25cm}}|@{\hspace{0.25cm}}
c@{\hspace{0.25cm}}c@{\hspace{0.25cm}}c@{\hspace{0.25cm}}c@{\hspace{0.25cm}}|}\hline
Dataset & $d_\sigma^{min}$ & $\Omega_{0m}^{min}$ & $w_0$ & $w_1$
\\\hline
Gold06 & 0.11 & 0.345 & -1.56 & 2.80 \\
Union & 0.17 & 0.300 & -1.26 & 1.25 \\
ESSENCE & 0.19 & 0.275 & -1.21 & 0.99 \\
SNLS1 & 0.42 & 0.290 & -1.04 & -0.26 \\
SDSS-II (SALT2) & 0.63 & 0.275 & -1.10 & 0.46 \\
SDSS-II (MLCS2K2) & 0.87 & 0.360 & -1.06 & 0.93 \\
Constitution & 1.77 & 0.315 & -0.88 & -1.32\\\hline
\end{tabular}}
\end{center}
\caption{The minimum $\sigma$-distances $d_\sigma^{min}$ to the
reference point $(w_0,w_1)=(-1.4,2)$, the values of $\omm$ at
which the minimum distance is attained, and the best fit
parameters $(w_0,w_1)$ at $\omm^{min}$ are displayed for each
dataset (see also Fig.~\ref{fig3}b). We omit the rows
corresponding to CMB+BAO data as the resulting $\sigma$-distance
is always $\gg1$ due to the dominance of the dark energy at early
times. The SDSS-II (MLCS2K2) data showed no minimum of $d_\sigma$
with respect to $\omm$ in the range $\omm \in [0.2,0.36]$. We thus
have simply displayed the lowest value of $d_\sigma$ in the
corresponding range of $\omm$.}\label{tabsep2}
\end{table}

The following comments can be made with respect to the results
shown in Figs.~\ref{fig3}a, \ref{fig3}b and in the corresponding
Tables \ref{tabsep1}, \ref{tabsep2}.
\begin{enumerate}
\item The consistency with $\Lambda$CDM of all datasets, except
Gold06  and SDSS-II when using the MLCS2K2 method, is maximized in
a narrow range of $\omm \in [0.26, 0.29]$ which also includes the
value of $\omm$ favored by standard rulers. On the other hand, the
consistency with the dynamical dark energy point $(w_0,
w_1)=(-1.4,2)$ is maximized over a wider range of $\omm$ ($\omm
\in [0.27, 0.35]$) thus decreasing the consistency among the
datasets in the context of dynamical dark energy. \item The
ranking sequence changes dramatically when the consistency with
the dynamical dark energy is considered as a reference point
instead of $\Lambda$CDM (Table \ref{tabsep2}). Essentially the
ranking is reversed! Thus, the choice of the consistency reference
point plays an important role in determining the ranking sequence
of the datasets (see also Fig.~3). \item The SDSS-II dataset
obtained with the MLCS2k2 fitter has some peculiar features
compared to other datasets. In particular it favors particularly
high values of $\omm$ ($\omm \simeq 0.4$) while for $\omm <0.3$
its consistency with \lcdm is significantly reduced to a level of
$3\sigma$ or larger ($d_\sigma > 3$). In addition, the trajectory
of its best fit parameter point as $\omm$ varies is perpendicular
to the corresponding trajectory of most other datasets (see Fig.
3).
\end{enumerate}

\begin{table}[!t]
\begin{center}
{\renewcommand{\arraystretch}{1.3}
\begin{tabular}{|c|c|c|}\hline Dataset & Prob. of Consistency &
Best Fit $\Omega_{0m}$  \\\hline
SNLS1 & 79\% & 0.26  \\
SDSS2-SALT2 & 68\% & 0.28  \\
SDSS2-MLCS2k2 & 52\% & 0.40  \\
ESSENCE & 30.4\% & 0.27  \\
Constitution & 12.6 \% & 0.29  \\
Union & 5.3 \% & 0.29  \\
Gold06 & 2.5\% & 0.34  \\\hline
\end{tabular}}
\end{center}
\caption{Consistency of SnIa datasets with $\Lambda$CDM according to the BND
statistic. Notice the high $\omm$ value favored by the SDSS2-MLCS2k2 dataset which is a few $\sigma$ above the value favored by other observations.}\label{tabBND}
\end{table}

An alternative ranking of the SnIa datasets according to their
consistency with $\Lambda$CDM can be made using the BND statistic
described briefly in section 2 and in more detail in Ref.
\cite{Perivolaropoulos:2008yc}. When applying the BND statistic to
find the consistency of a given dataset with $\Lambda$CDM, we find
the fraction of Monte Carlo datasets (generated from the best fit
$\Lambda$CDM model) that can mimic a BND crossing redshift $z_c$
(or crossing bin-size $N_c$) similar to that of the real data.
These Monte-Carlo datasets have $z_{mc}\leq z_c$ (or equivalently
$N_{mc}\geq N_c$) and their fraction represents a probability of
consistency of the given dataset with the model used to generate
the Monte-Carlo datasets (best fit $\Lambda$CDM).

\begin{figure}[!t]
\rotatebox{0}{\hspace{0cm}\resizebox{0.5\textwidth}{!}{\includegraphics{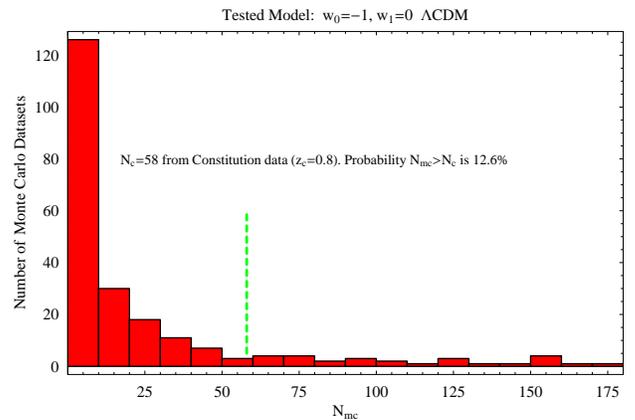}}}
{\hspace{0pt}\caption{The distribution of the crossing bin sizes
$N_{mc}$ of 222 Monte-Carlo datasets along with the crossing bin
size of the real Constitution data (thick dashed green
line).}\label{fig4}}
\end{figure}

An example of a distribution of crossing bin sizes $N_{mc}$ of
such Monte-Carlo datasets is shown in Fig.~\ref{fig4} for the case
of the Constitution dataset. In this case the best fit
$\Lambda$CDM model has $\omm=0.29$ and it is used to generate 222
Monte-Carlo realizations of the Constitution dataset as described
in Ref. \cite{Perivolaropoulos:2008yc}. The distribution of the
crossing bin sizes $N_{mc}$ of these datasets are shown in
Fig.~\ref{fig4} along with the crossing bin size of the real
Constitution data (thick dashed green line). In this case, the
probability of consistency of the Constitution dataset with the
Monte Carlo data of $\Lambda$CDM ($N_{mc}>N_c$) is $12.6\%$.

\begin{table}[!t]
\begin{center}
{\renewcommand{\arraystretch}{1.3}
\begin{tabular}{|@{\hspace{0.25cm}}c@{\hspace{0.25cm}}|@{\hspace{0.25cm}}c@{\hspace{0.25cm}}
c@{\hspace{0.25cm}}c@{\hspace{0.25cm}}c@{\hspace{0.25cm}}|@{\hspace{0.25cm}}c@{\hspace{0.25cm}}c@{\hspace{0.25cm}}|}\hline
Dataset & $d_\sigma^{min}$ & $\Omega_{0m}^{min}$ & $w_0$ & $w_1$ &
$w_0^{SR}$ & $w_1^{SR}$ \\\hline
SNLS1 & 0.003 & 0.280 & -1.04 & -0.10 & -1.07 & 0.08 \\
SDSS-II (SALT2) & 0.058 & 0.280 & -1.11 & 0.40 & -1.07 & 0.08 \\
ESSENCE & 0.087 & 0.275 & -1.21 & 0.99 & -1.12 & 0.38 \\
Constitution & 0.121 & 0.295 & -0.90 & -0.76 & -0.84 & -1.28 \\
Union & 0.681 & 0.280 & -1.24 & 1.44 & -1.07 & 0.08 \\
Gold06 & 1.976 & 0.280 & -1.38 & 2.75 & -1.07 & 0.08\\
{\scriptsize SDSS-II(MLCS2K2)} & 3.342 & 0.285 & -0.92 & 1.02 &
-1.00 & -0.28\\\hline
\end{tabular}}
\end{center}
\caption{Consistency with standard rulers. Minimum
$\sigma$-distance $d_{\sigma i}^{min}(\omm;(w_0,w_1)^{SR})$
between best fit parameters for each dataset, $(w_0,w_1)$, and
best fit for standard rulers, $(w_0,w_1)^{SR}$. $d_{\sigma
i}^{min}(\omm;(w_0,w_1)^{SR})$ is minimized at $\omm^{min}$ (see
Fig. 6). }\label{tabsep3}
\end{table}

Repeating the same process for all six datasets of Table
\ref{tabsnsets} we assign to each of them a probability of
consistency with $\Lambda$CDM which is shown in Table
\ref{tabBND}. Notice that the ranking sequence of consistency with
$\Lambda$CDM  obtained with the BND statistic is practically
equivalent with the corresponding ranking obtained with the
$\sigma$-distance statistic. This is reassuring for both ranking
approaches.

\begin{figure}[!t]
\vspace{.3cm}\epsfig{file=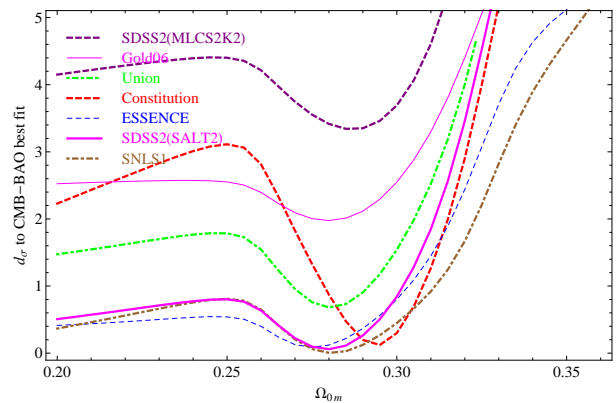,width=8cm}\caption{$\sigma$-distance
$d_{\sigma i} (\omm; (w_0, w_1)^{SR})$ between the reference point
$(w_0,w_1)^{SR}$ of standard rulers (using the WMAP5+SDSS5 data)
and the best fit of each dataset as a function of $\omm$ for the
datasets of Table \ref{tabsnsets}. Using the SDSS7 data, the
minimum distances are found in the range $\omm\in[0.29,0.31]$.}
\label{fig5}
\end{figure}

It is straightforward to apply the $\sigma$-distance statistic to
rank the SnIa datasets according to their consistency with
standard ruler CMB-BAO data. We simply use as a consistency
reference point the best fit point $(w_0,w_1)^{SR}$ for standard
rulers obtained as described in section 2 using the WMAP5+SDSS5
data. In this case, the location of the reference point
$(w_0,w_1)^{SR}$ in parameter space depends on $\omm$ but this
does not complicate the analysis. The $\sigma$-distance between
the reference point $(w_0,w_1)^{SR}$ and the best fit of each
dataset is shown in Fig.~\ref{fig5} as a function of $\omm$ for
the datasets of Table \ref{tabsnsets}. These distances are
minimized for values of $\omm$ that are different for each dataset
but they are all in the narrow range $\omm \in [0.27,0.3]$.

These minimum distances along with the corresponding value of
$\omm$ are shown in Table \ref{tabsep3} for each dataset (properly
ranked according to consistency with standard rulers). Notice that
the ranking sequence for the consistency with standard rulers is
practically identical to the ranking sequence of the consistency
with $\Lambda$CDM (Table \ref{tabsep1}) but differs from the
ranking sequence of the consistency with dynamical dark energy
(Table \ref{tabsep2}). This is an interesting feature of the data
in favor of $\Lambda$CDM.

Using the SDSS7 data the distance to the standard rulers best
fit for each dataset we find the minima shifted at slightly larger values
of $\omm$. We find that all of the minima lie in the range
$\omm\in[0.29,0.31]$ whereas the raking sequence of Table
\ref{tabsep3} is maintained.

As shown in Fig. 6 the SDSS-II dataset obtained with the MLCS2k2
fitter has reduced consistency not only with \lcdm (as noted
above) but also with standard rulers. Its distance from the
standard ruler best fit is at the level of $3-4\sigma$ ($d_\sigma
\simeq 3-4$). This is an additional indication of the peculiar
nature of this dataset. In contrast the SDSS-II dataset obtained
with the SALT2 fitter appears normal and consistent with \lcdm,
standard ruler and with the rest of the datasets (see Fig.
3).\footnote{In an effort to test the validity of our analysis for
the SDSS-II datasets we have reproduced successfully the $\chi^2$
contours of Ref. \cite{Kessler:2009ys} (Figs. 26e and 35e) in the simplified constant
$w$ parametrization considered there. Note that following Ref. \cite{Kessler:2009ys} we have only considered statistical errors.}

\section{Internal Consistency of SnIa Dataset Collections}
\begin{figure}[!t]
\hspace{-0.5cm}\epsfig{file=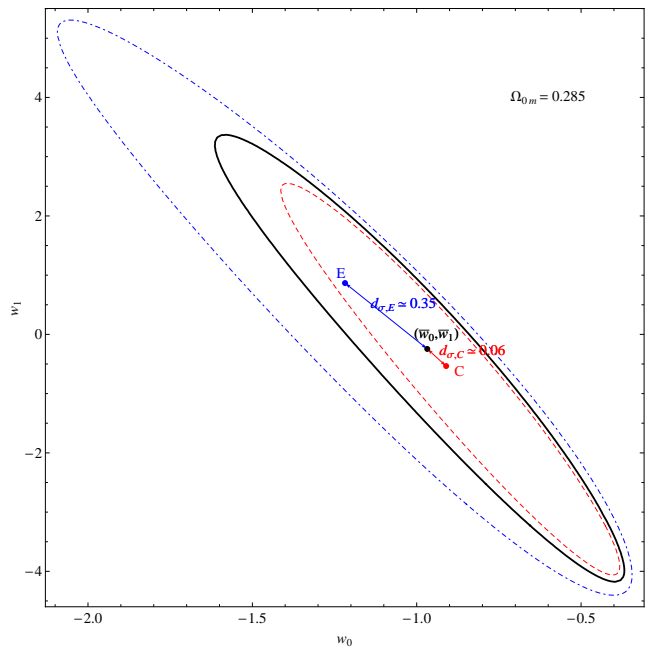,width=8.5cm}\caption{Construction
of ${\bar d}_{\sigma} (\omm; w_0, w_1)$ for a given point
$(w_0,w_1)$ in parameter space. The $\sigma$-distance between the
reference point and the best fit of each SnIa dataset in a
particular collection is evaluated using likelihood methods and
then the mean $\sigma$-distance is obtained using Eq.~(\ref{mtd}).
The figure illustrates the point $(\bar{w}_0,\bar{w}_1)$ where the
mean $\sigma$-distance $\bar{d}_\sigma$ (obtained by collecting
the Constitution and ESSENCE datasets) attains its minimum for a
prior $\omm=0.285$. The points $E$ and $C$ represent the best fit
parameters for the Constitution and ESSENCE datasets. Also in the
figure we show the $95.4\%$ contours for the Constitution (dashed)
and ESSENCE (dot-dashed) datasets, and the contour ${\bar
d}_{\sigma} (\omm; w_0, w_1)=2$ (thick, solid line).}\label{figg}
\end{figure}

In this section we test the internal consistency of various
collections of SnIa datasets using the mean $\sigma$-distance
statistic ${\bar d}_{\sigma} (\omm; w_0, w_1)$ defined by Eq.~
(\ref{mtd}). In Fig.~\ref{figg} we show an example demonstrating
the construction of ${\bar d}_{\sigma} (\omm; w_0, w_1)$ for a
given point $(w_0,w_1)$ in parameter space.

\begin{table}[!t]
\begin{center}
{\renewcommand{\arraystretch}{1.3}
\begin{tabular}{|@{\hspace{0.25cm}}c@{\hspace{0.25cm}}|@{\hspace{0.25cm}}c@{\hspace{0.25cm}}c@{\hspace{0.25cm}}|}\hline
Collection & $\langle\bar{d}_\sigma^{min}\rangle$ & Var
$\bar{d}_\sigma^{min}$\\\hline
ES & 0.03 & 0.0010 \\
CS & 0.13 & 0.0001 \\
CES & 0.20 & 0.0003 \\
CE & 0.21 & 0.0022 \\
CUES & 0.26 & 0.0043 \\
CUS & 0.28 & 0.0093 \\
CUE & 0.29 & 0.0147 \\
CU & 0.31 & 0.0392 \\
SG & 0.45 & 0.0194 \\
CUESG & 0.47 & 0.0163\\\hline
\end{tabular}}
\end{center}
\caption{Table of internal consistency of dataset collections. The
different datasets are labelled by C (Constitution), U (Union), E
(ESSENCE), S (SNLS1), and G (Gold06).
$\langle\bar{d}_\sigma^{min}\rangle$ and Var
$\bar{d}_\sigma^{min}$ stand for the average
$\bar{d}_\sigma^{min}$ over the range $\omm\in[0.20,0.36]$ and its
variance, respectively.}\label{tabcomb1}
\end{table}

\begin{table}[!t]
\begin{center}
{\renewcommand{\arraystretch}{1.3}
\begin{tabular}{|@{\hspace{0.25cm}}c@{\hspace{0.25cm}}|@{\hspace{0.25cm}}
c@{\hspace{0.25cm}}c@{\hspace{0.25cm}}|}\hline Collection &
$\bar{d}_{\sigma}^{min}$ & $\Omega_{0m}^{min}$\\\hline
ES & 0.12 & 0.265\\
CS & 0.16 & 0.280 \\
CES & 0.20 & 0.275\\
CE & 0.21 & 0.280 \\
CUES & 0.29 & 0.280\\
CUS & 0.29 & 0.280 \\
CU & 0.32 & 0.285 \\
CUE & 0.32 & 0.285\\
CUESG & 0.57 & 0.285\\
SG & 0.93 & 0.290\\\hline
\end{tabular}}
\end{center}
\caption{Minimum mean $\sigma$-distance
$\bar{d}_{\sigma}^{min}(\omm^{min};-1,0)$ from the best fit
parameters of various dataset collections, to the
$\Lambda$CDM parameter point.}\label{tabcomb3}
\end{table}

We minimize ${\bar d}_\sigma(\omm; w_0, w_1)$ with respect to
$(w_0,w_1)$ and find the parameter point $({\bar w_0}, {\bar
w_1})$ of maximum consistency with the given dataset collection as
well as the minimum mean $\sigma$-distance ${\bar
d}_\sigma^{min}(\omm;{\bar w}_0, {\bar w}_1)$ indicating the level
of consistency. This minimized distance is shown in Fig.~8 as a
function of $\omm$ for various dataset collections. Clearly,
${\bar d}_\sigma^{min}(\omm;{\bar w}_0, {\bar w}_1)$ is weakly
dependent on $\omm$ but depends sensitively on the dataset
collection considered.

Marginalizing ${\bar d}_\sigma^{min}(\omm;{\bar w}_0, {\bar w}_1)$
with respect to $\omm$ in the range $\omm \in [0.2,0.36]$ we find
$\langle \bar{d}_\sigma^{min}\rangle$ and its variance which may
be directly used as a measure of the internal consistency of a given
dataset collection.

\begin{figure}[!t]
\begin{center}
$\begin{array}{@{\hspace{-0.3cm}}c@{\hspace{0.10cm}}c}
\epsfig{file=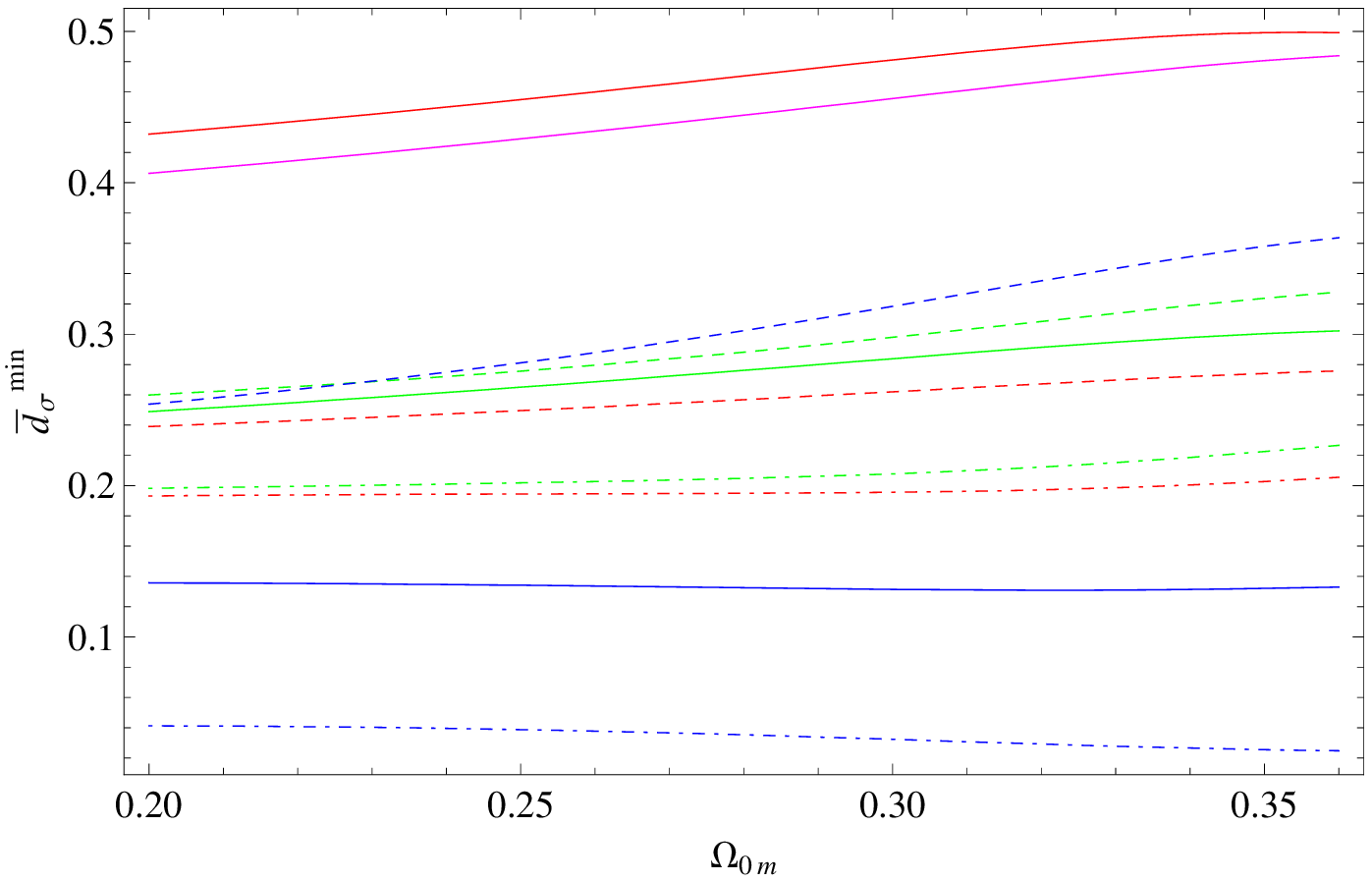,width=7.3cm} &
\epsfig{file=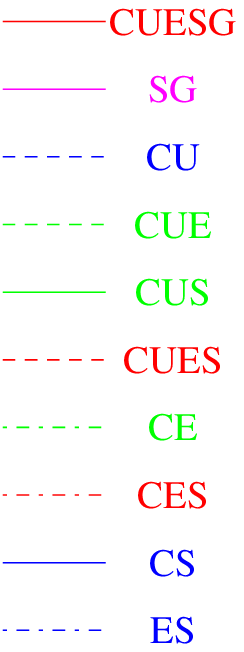,bb=0 -50 69 192,width=1.3cm}\\
\end{array}$
\end{center}
\caption{Minimum mean $\sigma$-distance
$\bar{d}_\sigma^{min}(\omm;w_0,w_1)$ for the dataset collections
considered in Table \ref{tabcomb1}. The sequence of the legend
corresponds to the sequence of lines in the plot.}\label{fig6}
\end{figure}

To summarize, the derivation and use of our statistic involves the following steps:
\begin{enumerate}
\item Minimize the mean $\sigma$-distance defined by Eq.~
(\ref{mtd}), with respect to $(w_0,w_1)$. \item   Marginalize the
resulting minimum mean $\sigma$-distance over $\Omega_{0m}$. This
step is not important due to the weak dependence of the minimum
$\sigma$-distance on $\Omega_{0m}$ (see Fig.~8). The evaluation of
the (small) variance with respect to $\Omega_{0m}$  is only used
to demonstrate the weak dependence of the mean $\sigma$-distance
on  $\Omega_{0m}$  and plays no further role in the analysis.
\item   Use the  marginalized (over $\Omega_{0m}$) minimum (with
respect to $(w_0,w_1)$) mean $\sigma$-distance as a measure of the
internal consistency of the dataset collection. The smaller the
minimum mean $\sigma$-distance is, the 'closer' together are the
best fit parameter values in parameter space.
\end{enumerate}
The values of $\langle
\bar{d}_\sigma^{min}\rangle$ and their variances for various
dataset collections (properly ranked) are shown in Table
\ref{tabcomb1}.
The following comments can be made with respect to the results
shown in Table \ref{tabcomb1}: \begin{itemize} \item All dataset
collections considered are mutually consistent in the sense that the mean
$\sigma$-distance between the best fits of each dataset and the
point of maximum consistency is less than 1 ($1\sigma$). \item
Collections including the Gold06 dataset have significantly less
internal consistency than other dataset collections. \item Maximum
internal consistency is achieved for the ESSENCE-SNLS1 collection
($\langle d_\sigma\rangle=0.03$) and for the Constitution-SNLS1
collection ($\langle d_\sigma\rangle=0.13$) which are collections
that also maximize consistency with $\Lambda$CDM and CMB-BAO
standard rulers (see below).
\end{itemize}

In order to rank the consistency of {\it collections} of datasets
with $\Lambda$CDM and with standard rulers we consider the mean
$\sigma$-distances from $\Lambda$CDM ${\bar d}_\sigma(\omm; -1,
0)$ and from the standard ruler best fit ${\bar d}_\sigma(\omm;
w_0^{SR}, w_1^{SR})$ (for the standard ruler best fit
$(w_0^{SR},w_1^{SR})$). These are shown in Fig.~9 as functions of
$\omm$. We then minimize these distances with respect to $\omm$
and find the corresponding ${\bar d}_\sigma^{min}$ distances.
These may now be used to rank the dataset collections with respect
to their consistency with either $\Lambda$CDM or with standard
rulers. These rankings are shown in Table \ref{tabcomb3} (with
respect to $\Lambda$CDM $(-1,0)$) and in Table \ref{tabcomb2}
(with respect to standard ruler best fit $(w_0^{SR},w_1^{SR})$).

\begin{figure*}[!t]
\centering
\begin{center}
$\begin{array}{@{\hspace{-0.20in}}c@{\hspace{0.0in}}c}
\multicolumn{1}{l}{\mbox{}} & \multicolumn{1}{l}{\mbox{}} \\
[-0.2in]\epsfxsize=3.3in \epsffile{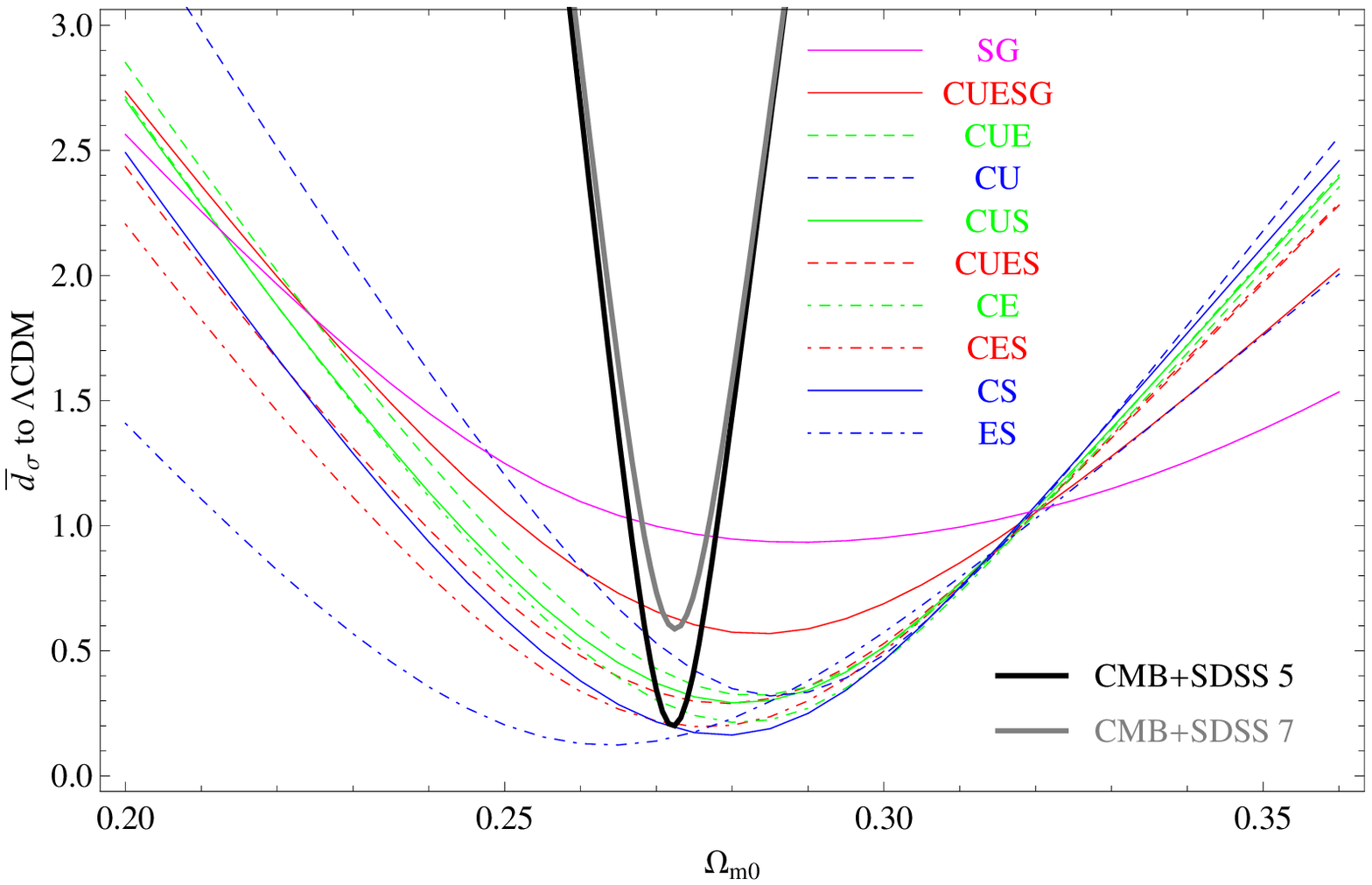} & \epsfxsize=3.3in
\hspace{0.20in}\epsffile{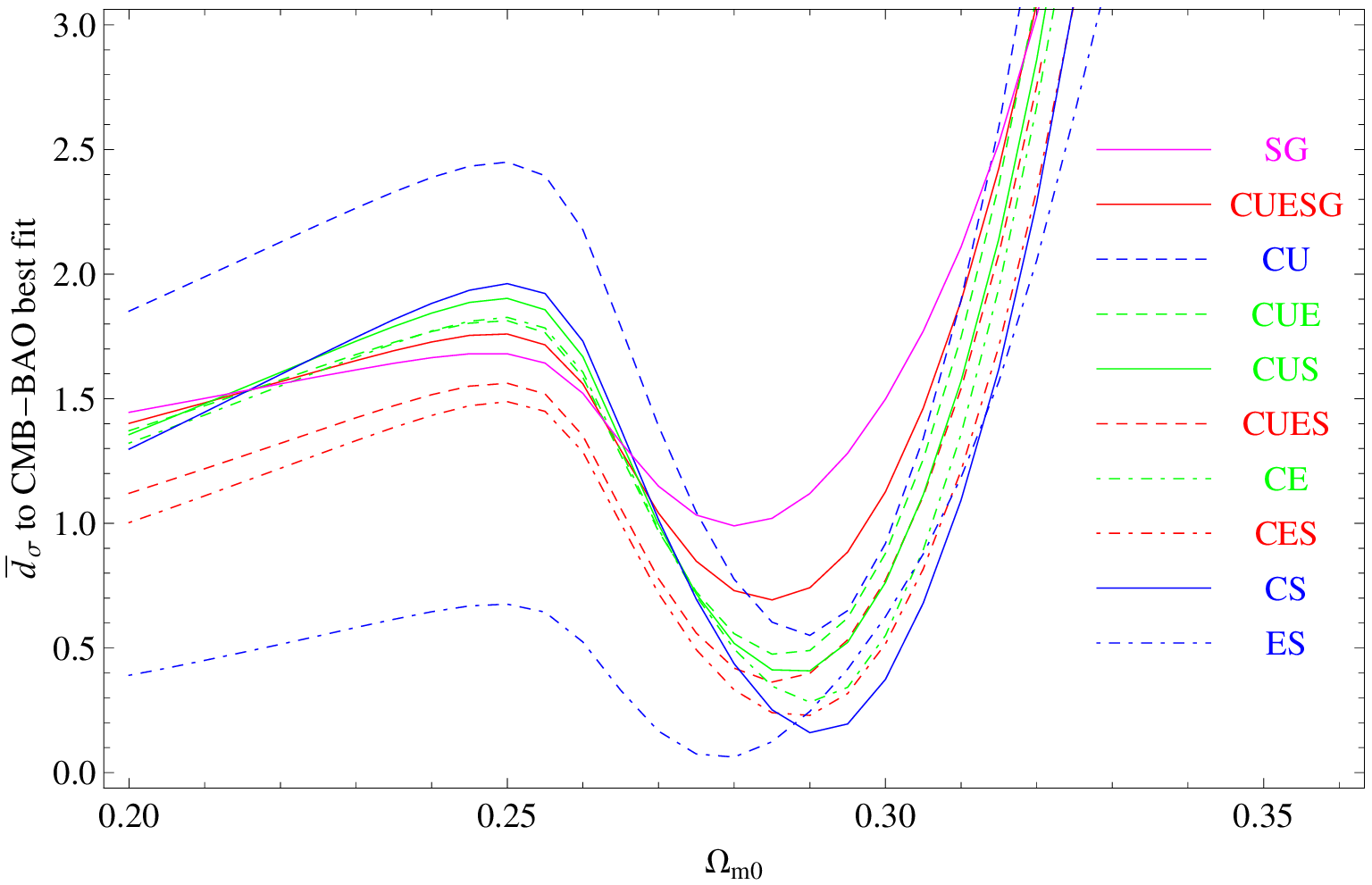} \\
\end{array}$
\end{center}
\vspace{0.0cm} \caption{\small a: $\sigma$-distances
$\bar{d}_\sigma(\omm;-1,0)$ of the $\Lambda$CDM point for the
dataset collections in Table \ref{tabcomb1}. The minimum values of
$\bar{d}_\sigma(\omm;-1,0)$ and the corresponding values of $\omm$
are collected in Table \ref{tabcomb3}. Also shown are the
$\sigma$-distances between the standard ruler best fit (SDDS5 and
SDSS7) and $\Lambda$CDM (black and grey solid lines). b:
$\sigma$-distances $\bar{d}_\sigma(\omm;(w_0,w_1)^{SR})$ to the
standard ruler best fit (WMAP5+SDSS5). The minimum values of
$\bar{d}_\sigma(\omm;(w_0,w_1)^{SR})$ and the corresponding values
of $\omm$ are listed in Table \ref{tabcomb2}.}
\end{figure*}

The following comments can be made with respect to the results
shown in Tables \ref{tabcomb3} and \ref{tabcomb2}:
\begin{itemize} \item The ranking sequences of dataset collections
with respect to a. consistency with $\Lambda$CDM, b. consistency
with standard rulers and c. internal consistency (Table
\ref{tabcomb1}) are practically identical. This is an interesting
result in view of the fact that the three ranking criteria are
independent of each other. Criteria a. and c. rank the quality of
the datasets while criterion b. ranks the consistency with a given
cosmological model. Thus, this coincidence may be viewed as
evidence supporting the $\Lambda$CDM model since other parameter
reference points would in general tend to spoil this ranking (see
Table \ref{tabsep2}). \item More consistent collections (with any
of the criteria) are those including the SNLS1 dataset while less
consistent are those including the Gold06 dataset. \item Maximum
consistency for all dataset collections with either $\Lambda$CDM
or with standard rulers is achieved for a narrow range of $\omm$
($\omm \in [0.27,0.29]$). This is an indication of the robustness
of the criteria used.
\end{itemize}

\begin{table}[!t]
\begin{center}
{\renewcommand{\arraystretch}{1.3}
\begin{tabular}{|@{\hspace{0.25cm}}c@{\hspace{0.25cm}}|@{\hspace{0.25cm}}
c@{\hspace{0.25cm}}c@{\hspace{0.25cm}}c@{\hspace{0.25cm}}c@{\hspace{0.25cm}}|}\hline
Collection & $\bar{d}_{\sigma}^{min}$ & $\Omega_{0m}^{min}$ &
$w_0^{SR}$ & $w_1^{SR}$\\\hline
ES & 0.06 & 0.280 & -1.07 & 0.08 \\
CS & 0.16 & 0.290 & -0.93 & -0.73 \\
CES & 0.23 & 0.290 & -0.93 & -0.73 \\
CE & 0.28 & 0.290 & -0.93 & -0.73 \\
CUES & 0.36 & 0.285 & -1.00 & -0.28 \\
CUS & 0.41 & 0.290 & -0.93 & -0.73 \\
CUE & 0.48 & 0.285 & -1.00 & -0.28 \\
CU & 0.55 & 0.290 & -0.93 & -0.73 \\
CUESG & 0.69 & 0.285 & -1.00 & -0.28 \\
SG & 0.99 & 0.280 & -1.07 & 0.08\\\hline
\end{tabular}}
\end{center}
\caption{Minimum mean $\sigma$-distance
$\bar{d}_{\sigma}^{min}(\omm^{min};(w_0,w_1)^{SR})$ from the best
fit parameters of each dataset in a particular collection to the
standard ruler best fit parameters
$(w_0,w_1)^{SR}$.}\label{tabcomb2}
\end{table}

\section{Conclusions}
The main conclusions of our analysis comparing the six most recent
SnIa datasets in the context of the CPL parametrization may be
summarized as follows: \begin{itemize}\item  All datasets can be
made consistent with $\Lambda$CDM and with standard rulers at a
level of $95.4\%$ ($2\sigma$) or less for certain prior values of
the matter density $\omm$ in the range $\omm \in [0.25,0.35]$.
\item The Gold06 and the SDSS-II standardized with MLCS2k2
datasets have the minimum consistency with both $\Lambda$CDM and
with standard rulers while SNLS1 is the most consistent dataset
with both $\Lambda$CDM and standard rulers. This may be related to
the fact that Gold06 is highly inhomogeneous and includes a few
outliers \cite{Nesseris:2006ey} while SNLS1 is the most
homogeneous dataset. These consistency features are inherited (at
a reduced degree) to other datasets of Table 1 that include parts
of SNLS1 (e.g. Constitution or ESSENCE) or Gold06 (i.e. Union).
\item The best fit of the Gold06 dataset (and of the Union that
includes part of it) corresponds to a universe that
super-accelerates at present ($w(z=0)=w_0<-1$) after having
crossed the phantom divide line (PDL) $w=-1$ at a recent redshift
($w_1>0$) (see Fig.~3). On the other hand the best fit of the
Constitution dataset gives a reversed behavior: no
super-acceleration at present ($w_0 >-1$) but for $\omm>0.25$ it
crosses the PDL at a recent redshift leading $w(z)<-1$ in the
recent past since $w_1<0$ (see Fig.~3). This behavior has been
discussed in some detail in Ref. \cite{Shafieloo:2009ti}.  We
stress that this is only the behavior of the best fit and should
not be taken as a statistically significant trend of the cosmic
history indicated by the datasets. \item All six datasets are
statistically mutually consistent. However, this consistency is
somewhat reduced for dataset collections that include the Gold06
dataset (see Fig.~8 and Table \ref{tabcomb1}). This is consistent
with previous analyses \cite{Nesseris:2006ey} that pointed out
non-uniformity systematics in the Gold06 dataset. \item The
SDSS-II dataset obtained with the MLCS2k2 fitter has reduced
consistency with both \lcdm and with standard rulers in contrast
with the same dataset standardized with the SALT2 fitter which
appears similar with the other datasets and consistent with LCDM.
We have verified that this is not a general deficiency of the
MLCS2k2 because other datasets using the same fitter (e.g. ESSENCE
or the MLCS version of the Constitution dataset) appear fairly
normal and consistent (see Fig. 3). Therefore we have not been
able to trace the origin of the peculiar nature of the SDSS-II
MLCS2k2 dataset.
\end{itemize}

It is interesting that despite the improvement of standard ruler
and standard candle data quality during the last decade the
consistency of $\Lambda$CDM with data has not decreased despite
the fact that $\Lambda$CDM is a simple, specific and well defined
model which appears as a measure-zero point in all generalized
models. On the contrary its consistency seems to be improving with
time as new and more accurate data appear. For example, the
Constitution SnIa dataset which is a very recent compilation with a
drastic improvement on the crucial nearby SnIa sample, is also one
of the most consistent datasets with both $\Lambda$CDM and
standard rulers.

Despite of its excellent consistency with both SnIa standard candles
and CMB-BAO standard rulers, $\Lambda$CDM has to face potential
challenges from other cosmological data
\cite{Perivolaropoulos:2008ud} (e.g. large scale velocity flows,
galaxy and cluster halo profiles, peculiar features of CMB maps
etc.) which may lead the quest for the properties of dark energy
to interesting surprises in the near future. Such surprises may
also come from future standard candle observations or standard
ruler CMB experiments (e.g. Planck \cite{planck}) which are
expected to significantly improve the accuracy of the constraints
discussed in the present study.

{\bf Numerical Analysis Files:} The mathematica files and datasets
used for the production of the figures may be downloaded from
http://leandros.physics.uoi.gr/datacomp/

We have tested that these files reproduce the results of the
original dataset papers
\cite{Hicken:2009dk,Kowalski:2008ez,Davis:2007na,Riess:2006fw,Astier:2005qq,Kessler:2009ys}
in the special case of constant $w$ considered there.

\section*{Acknowledgements}
We thank R.~Kessler for useful comments.
This work was supported by the European Research and Training
Network MRTN-CT-2006 035863-1 (UniverseNet). S.~N. also
acknowledges support by the Niels Bohr International Academy and
the Danish Research Council under FNU Grant No. 272-08-0285.

\end{document}